\documentclass[11pt]{article}

\usepackage[letterpaper,margin=1in]{geometry}
\usepackage{amsmath}
\usepackage{amssymb}
\usepackage{graphicx}
\usepackage[authoryear,round]{natbib}
\usepackage{booktabs}
\usepackage[hidelinks]{hyperref}

\providecommand{\doi}[1]{doi:\href{https://doi.org/#1}{#1}}

\newcommand{\Teff}{T_\mathrm{eff}}
\newcommand{\Tcore}{T_\mathrm{core}}
\newcommand{\TB}{T_B}
\newcommand{\DKL}{D_\mathrm{KL}}
\newcommand{\fk}{f_\kappa}
\newcommand{\Mxw}{f_\mathrm{Mxw}}
\newcommand{\Kn}{\mathrm{Kn}}
\newcommand{\dIS}{d_\mathrm{IS}}

\begin{document}

\title{The Diagnostic Disagreement and the Closure Failure of the Quiet-Sun
Corona Are the Same Relative Entropy}

\author{Victor Edmonds\thanks{\texttt{vedmonds@finalstopconsulting.com}}\\[3pt]
\normalsize Final Stop Consulting LLC, Raleigh, NC, USA\\
\normalsize Ronin Institute for Independent Scholarship 2.0, Sacramento, CA, USA\\}
\date{}

\maketitle

\begin{abstract}
In the weakly collisional quiet-Sun corona the electron distribution departs
from Maxwellian, and ``the electron temperature'' is not single-valued: two
standard diagnostics differ by a stable factor $R \approx 2.4$ across the
solar cycle. We show this departure is a single relative entropy whose two
orthogonal components are the EUV--radio diagnostic disagreement and the
failure of the local conductive closure. By the generalized
Pythagorean identity for Bregman divergences, the relative entropy of the
$\kappa \approx 2.5$ distribution from its radio-diagnostic Maxwellian
projection ($1.20$ nats) partitions with no cross-term into two components. The
first is the energy-matched closure deficit ($0.32$ nats), the information
discarded by the local Spitzer--H\"{a}rm closure \citep{edmonds2026c}, which
itself has no convergent form across $\kappa \in [2,3]$. The second is the inter-projection
diagnostic gap ($0.876$ nats), what the EUV and radio diagnostics disagree by
and a closed-form function of $R$ within the shape envelope. Closure failure and
diagnostic disagreement are therefore the same relative entropy in two
orthogonal coordinates: together they account for the corona's full departure from
Maxwellian. The physical heat flux that survives is non-local and tail-dominated.
Any mechanism that sets this $\kappa$ must act on the suprathermal tail,
operate non-locally, and fix $\kappa$ to the observed band; of the standalone candidates only velocity filtration
\citep{scudder1992a} meets the first two,
and whether it, or anything, fixes $\kappa$ is the open problem we leave
posed.
\end{abstract}


\section{Introduction}
\label{sec:intro}

The quiet-Sun corona is weakly collisional, and a weakly collisional plasma
need not be Maxwellian. At a base electron density $n_e \sim 10^{8}\,
\mathrm{cm}^{-3}$ and $T \sim 10^{6}\,\mathrm{K}$ the bulk Knudsen number is
$\Kn \sim 0.01$--$0.1$, and because the Coulomb mean free path scales as
$v^4$, the suprathermal electrons that carry conduction are effectively
collisionless even where the bulk is marginally collisional
\citep{shoub1983,landipantellini2001}. The electron velocity distribution
$f(\mathbf{v})$ is, on the evidence of two independent diagnostics, a
$\kappa$ distribution with $\kappa \approx 2.5$ \citep{edmonds2026a}. In such a
plasma ``the electron temperature'' is not single-valued. Each diagnostic
returns the temperature of the Maxwellian that matches the particular moment
of $f(\mathbf{v})$ its kernel weights, and different kernels return different
temperatures: the EUV ionization temperature $\Teff \approx 1.5\,\mathrm{MK}$
exceeds the radio brightness temperature $\TB \approx 0.6\,\mathrm{MK}$ by a
factor $R \equiv \Teff/\TB \approx 2.4$, stable across the 2004--2011 solar
cycle to within yearly variations below $10\%$
\citep{mercierchambe2015,vocks2018,zhang2022}.

The relative entropy of the
$\kappa \approx 2.5$ distribution from the Maxwellian its radio diagnostic
returns is $1.20$ nats (natural-log units of relative entropy), and by the
generalized Pythagorean identity for
divergences on exponential families it splits, with no cross-term, into two
pieces, each already known under another name:
\begin{equation}
\DKL(\fk \,\|\, \Mxw(\Tcore))
= \underbrace{\DKL(\fk \,\|\, \Mxw(\Teff))}_{0.32~\mathrm{nats}}
+ \underbrace{\DKL(\Mxw(\Teff) \,\|\, \Mxw(\Tcore))}_{0.876~\mathrm{nats}} .
\label{eq:partition-intro}
\end{equation}
The first piece, $0.32$ nats, is the energy-matched closure deficit: the
information the local Spitzer--H\"{a}rm conductive closure discards in
representing this distribution by a single Maxwellian \citep[derived in][]{edmonds2026c}.
The second, $0.876$ nats, is the diagnostic gap: the relative entropy between
the two Maxwellians the EUV and radio diagnostics return, a closed-form function
of $R$ within the shape envelope. Closure failure and diagnostic disagreement are therefore
not two problems but two orthogonal coordinates of one relative entropy, and
the partition accounts for the corona's departure from Maxwellian without remainder.

The decomposition is an identification of results already derived, not a new
claim: both components are computed in closed form (the diagnostic gap here, the
closure deficit in \citealt{edmonds2026c}), and their orthogonality is the
Pythagorean identity for the I-projection onto the Maxwellian family. What the
partition does not fix is the shape parameter itself. The flux that carries
the departure is non-local and tail-dominated; the closure that the deficit
measures does not exist as a local quantity across $\kappa \in [2,3]$. The
paper closes by reading three constraints off the structure of the entropy ---
that any generator of this $\kappa$ act on the tail, operate non-locally, and
fix $\kappa$ to the band --- showing that the candidate mechanisms reduce to the one whose
definition is a non-local transport statement,
and posing, without answering, whether that mechanism fixes $\kappa$.

\section{The diagnostic reading}
\label{sec:diagnostic}

Each coronal electron-temperature diagnostic returns a scalar by projecting
the full velocity distribution $f(\mathbf{v})$ onto the one-parameter family
of Maxwellians, replacing $f$ by the Maxwellian whose kernel-matched parameter
reproduces a specified property of the diagnostic kernel. Two standard
procedures correspond to two structurally inequivalent projection rules.

\subsection{Two projections onto the Maxwellian family}
\label{sec:projection}

Let $K_D(\mathbf{v})$ be a diagnostic kernel. For a measured observable of the
form $D[f] = \int K_D(\mathbf{v})\,f(\mathbf{v})\,d^3v$, the
\emph{moment-matched} temperature is the parameter of the Maxwellian whose
kernel-integrated observable reproduces $D[f]$,
\begin{equation}
\int K_D\,\Mxw(\mathbf{v}; T_D)\,d^3v = \int K_D\,f(\mathbf{v})\,d^3v .
\label{eq:moment-match}
\end{equation}
When the kernel coincides with the sufficient statistic of the family --- as
it does for the energy moment, $K_D \propto E$ --- the moment-matched member
is the information projection (I-projection) of $f$ onto the family in the
sense of \citet{csiszar1975}, the unique member minimizing $\DKL(f \,\|\,
\cdot)$.\footnote{Minimizing $\DKL(f\,\|\,\cdot)$ over an exponential family is,
strictly, the reverse $I$-projection (the $m$-projection of information
geometry); we keep the term ``$I$-projection'' for brevity. Identity
(\ref{eq:pythagoras}) is the generalized Pythagorean relation this projection
satisfies, and holds exactly.} The EUV ionization diagnostic implements this projection
(Section~\ref{sec:euv}).

When instead the diagnostic is a scattering process whose observable is the
ratio of an emissivity $j_\nu[f]$ to an absorption coefficient $\alpha_\nu[f]$,
the returned \emph{source-function} temperature is
\begin{equation}
T_D^{(\mathrm{SF})} = \frac{j_\nu[f]}{\alpha_\nu[f]}\,\frac{c^2}{2k\nu^2} ,
\label{eq:source-func}
\end{equation}
the temperature at which a Maxwellian plasma would produce the observed ratio
in the Rayleigh--Jeans limit. For a Maxwellian $f$ this collapses to the
plasma temperature by Kirchhoff's law; for a non-Maxwellian $f$ the numerator
and denominator are separate moments and their ratio is, in general, the
moment-matched temperature of neither. Thermal free-free bremsstrahlung in the
corona is this case (Section~\ref{sec:radio}). Both rules are projections onto
the same one-parameter family, and their outputs need not agree when $f$
departs from Maxwellian.

\subsection{The EUV side: moment-matching under ionization}
\label{sec:euv}

The dominant EUV line in the quiet-Sun comparison (Fe\,\textsc{xii} 195\,\AA;
\citealt{mercierchambe2015}) is populated through collisional ionization of
Fe\,\textsc{xi}, whose rate coefficient in the Bethe form is
$\sigma_\mathrm{ion}(v) \propto (1/E)\ln(E/\chi)$ above threshold $\chi$. For a
$\kappa$ distribution in the mean-energy (Dzif\v{c}\'{a}kov\'{a}--Dud\'{\i}k,
hereafter D\&D; \citealt{dd2013}) convention, normalized so that
$\langle E\rangle = \tfrac{3}{2}k\Teff$ for all $\kappa > 3/2$, the
moment-matching projection (\ref{eq:moment-match}) returns
\begin{equation}
T_\mathrm{EUV}^{(\mathrm{MM})} = \Teff\,[\,1 + \varepsilon_\mathrm{EUV}\,] ,
\label{eq:T-EUV}
\end{equation}
where $\varepsilon_\mathrm{EUV}$ measures the departure of the Bethe-weighted
ionization moment from the energy moment. \citet{dudik2014} establish that
strong Fe\,\textsc{ix}--\textsc{xiii} line intensity ratios are largely
$\kappa$-insensitive across $\kappa = 2$--$25$, a residual variation we read as
$\lesssim 20\%$; $\varepsilon_\mathrm{EUV}$ is bounded by this observational
envelope. The rigorous atomic-data evaluation of
Section~\ref{sec:numerical}, using the matched $\kappa$-resolved and Maxwellian
ion-equilibrium tables of \citet{dz23} on the CHIANTI v10.1 basis, returns a
single definite $\varepsilon_\mathrm{EUV}$ inside this bracket.

\subsection{The radio side: a known identity}
\label{sec:radio}

The radio brightness temperature in the Rayleigh--Jeans regime is the
source-function projection (\ref{eq:source-func}) with $j_\nu$ and $\alpha_\nu$
the emissivity and absorption coefficient of thermal free-free
bremsstrahlung. The brightness temperature equals this source function in the
optically thick limit ($\tau \gg 1$), which the quiet corona satisfies at
150--450\,MHz \citep{mercierchambe2015,zhang2022}; at finite optical depth
$\TB$ would fall below it and inflate $R$. In the kinetic limit
$h\nu \ll k\Teff$, the emissivity is set
by $\langle 1/v\rangle_f$ and the absorption coefficient, by microscopic
reversibility, collapses to a boundary term controlled by the zero-velocity
density $f_v(0)$. For a $\kappa$ distribution in the D\&D convention,
\citet{fk2014} (their Eq.~48) give the resulting source function in closed
form:
\begin{equation}
\TB = \frac{\kappa - 3/2}{\kappa}\,\Teff \equiv \Tcore ,
\label{eq:TB-is-Tcore}
\end{equation}
where $\Tcore$ is the core temperature scalar (the most-probable-energy
Maxwellian temperature of \citealt{oka2013}). We adopt
(\ref{eq:TB-is-Tcore}) as input; it is algebraic, independent of frequency,
density, and Coulomb logarithm in the slowly-varying Gaunt approximation, and
recovers the Maxwellian limit as $\kappa \to \infty$. Any residual finite-$\kappa$
departure of the source function from $\Tcore$ beyond this approximation would
shift $\TB$ to a different member of the Maxwellian family; because the partition
of Section~\ref{sec:unification} holds for any member, such a correction changes
only the numerical split between the two components, not the orthogonal structure.
The diagnostic ratio is
therefore a closed-form function of $\kappa$ alone,
\begin{equation}
R \equiv \frac{T_\mathrm{EUV}^{(\mathrm{MM})}}{\TB}
= \frac{\kappa}{\kappa - 3/2}\,[\,1 + \varepsilon_\mathrm{EUV}\,] ,
\label{eq:R-closed-form}
\end{equation}
its only approximation the bounded EUV-side correction
$\varepsilon_\mathrm{EUV}$.

\subsection{Numerical evaluation}
\label{sec:numerical}

We fix $\Teff = 1.5\,\mathrm{MK}$ and $\kappa = 2.5$ from the joint cross-fit
of \citet{edmonds2026a}, in which spectroscopic EUV line ratios, radio
brightness at 150--450\,MHz, and the hydrostatic scale-height density profile
intersect at $\kappa \approx 2.5$; we take this value as input and do not
re-derive it. Both diagnostics integrate along the line of sight, where a
multithermal superposition of Maxwellians can mimic a single $\kappa$
distribution; disentangling that degeneracy from a genuine non-Maxwellian
departure is the work of the cross-fit and DEM analysis we take as input
\citep{edmonds2026a,edmonds2026c}; in brief, the radio side breaks it: free-free
bremsstrahlung carries no ionization threshold and is set by the distribution
core, so a multithermal mix tuned to the EUV line ratios predicts a radio
brightness temperature it cannot match, and the diagnostics intersect only near
a genuine $\kappa \approx 2.5$. The radio side gives $\TB = \Tcore = 0.60\,
\mathrm{MK}$ at the order of the approximations under which
(\ref{eq:TB-is-Tcore}) is derived. The EUV side, with
the Bethe kernel, gives $\varepsilon_\mathrm{EUV} = 0.013$ as a baseline;
alternative cross-section shapes bracket $T_\mathrm{EUV} \in 123$--$154\,
\mathrm{eV}$, i.e. $\varepsilon_\mathrm{EUV} \in [-0.05, +0.19]$, all inside
the \citet{dudik2014} envelope by construction. The line-ratio inversion we
compute on the \citet{dz23} tables returns $T_M = 1.47\,\mathrm{MK}$ for Fe\,\textsc{xii}/
Fe\,\textsc{xiii} ($\varepsilon_\mathrm{EUV} = -0.020$, $R = 2.45$) and
$T_M = 1.35\,\mathrm{MK}$ for Fe\,\textsc{xi}/Fe\,\textsc{xii}
($\varepsilon_\mathrm{EUV} = -0.10$, $R = 2.25$). The Fe\,\textsc{xii}/
Fe\,\textsc{xiii} value sits centrally within the \citet{mercierchambe2015}
yearly range $R = 2.31$--$2.53$; the Fe\,\textsc{xi}/Fe\,\textsc{xii} value lies
just below its lower edge and just outside the shape bracket above, while
remaining inside the broader \citet{dudik2014} envelope. The closed-form ideal $R_0 \equiv
\kappa/(\kappa - 3/2) = 2.500$ at $\kappa = 2.5$.

\subsection{The diagnostic gap as relative entropy}
\label{sec:gap}

For a $\kappa$ distribution in the D\&D convention the Kullback--Leibler
divergence from a Maxwellian at temperature $T$ admits a closed form in the
log-gamma and digamma functions,
\begin{align}
\DKL(\fk \,\|\, \Mxw(T))
&= \ln\frac{\Gamma(\kappa+1)}{\Gamma(\kappa-1/2)}
 + \frac{3}{2}\ln\frac{T}{T_0} \nonumber\\
&\quad - (\kappa+1)\,[\psi(\kappa+1) - \psi(\kappa-1/2)]
 + \frac{3}{2}\frac{\Teff}{T} ,
\label{eq:DKL-closed-form}
\end{align}
with $T_0 = (\kappa - 3/2)\,\Teff$ and $\psi(x) = d\ln\Gamma(x)/dx$ the
digamma function. At
$\kappa = 2.5$, $\Teff = 1.5\,\mathrm{MK}$,
\begin{align}
\DKL(\fk \,\|\, \Mxw(\Teff)) &= 0.320~\mathrm{nats}, \label{eq:dkl-eff}\\
\DKL(\fk \,\|\, \Mxw(\Tcore)) &= 1.195~\mathrm{nats}. \label{eq:dkl-core}
\end{align}
Their difference follows from (\ref{eq:DKL-closed-form}) directly: the
$\Gamma$-ratio and digamma terms cancel, and
\begin{equation}
\Delta\DKL \equiv \DKL(\cdot\|\Mxw(\Tcore)) - \DKL(\cdot\|\Mxw(\Teff))
= \tfrac{3}{2}\,[\,R_0 - \ln R_0 - 1\,] ,
\label{eq:gap-identity}
\end{equation}
which at $\kappa = 2.5$ evaluates to $0.876$ nats. The functional
$\dIS(x,y) \equiv x/y - \ln(x/y) - 1$ is the Itakura--Saito distance, the
canonical Bregman divergence generated by $\phi(z) = -\ln z$
\citep{banerjee2005}; with $x = \Teff$, $y = \Tcore$, identity
(\ref{eq:gap-identity}) reads $\Delta\DKL = \tfrac{3}{2}\,\dIS(\Teff,\Tcore)$. It
appears because the gap is the relative entropy between two Maxwellians
differing only in temperature, and the Kullback--Leibler divergence between two
zero-mean Gaussians differing only in scalar variance is, per dimension,
one-half the Itakura--Saito distance of those variances; the factor
$\tfrac{3}{2}$ in (\ref{eq:gap-identity}) collects the three velocity-space
dimensions at that one-half each. The
Itakura--Saito distance is scale-invariant,
$\dIS(cx,cy) = \dIS(x,y)$, so the gap depends only on the ratio $R$ and not on
the absolute coronal temperature.
The $0.876$-nat gap is thus the relative entropy between the two diagnostic
Maxwellians themselves, and the observed $R \approx 2.4$ is a direct
measurement of it within the shape envelope. Over the \citet{mercierchambe2015}
solar-cycle range $R \in [2.31, 2.53]$, the gap runs from $0.71$ to $0.90$ nats.
The cancellation is structural:
$\Mxw(\Teff)$ is the I-projection of $\fk$, so the two components are orthogonal
(Section~\ref{sec:unification}).

\section{A single diagnostic is predicted to see a Maxwellian}
\label{sec:appearance}

\citet{lorincik2020} report that quiet-Sun Hinode/EIS spectra are consistent
with a Maxwellian, with only active-region loops and moss requiring
$\kappa \lesssim 3$. This is not in tension with
$\kappa = 2.5$; it is a consequence of the projection structure of
Section~\ref{sec:diagnostic}.

Under the D\&D convention $\langle E\rangle$ is held fixed across $\kappa$ by
construction, and the EUV diagnostic projects $\fk$ onto the Maxwellian family
under a Bethe-weighted ionization kernel whose strong-line intensity ratios
are largely $\kappa$-insensitive across $\kappa = 2$--$25$ \citep{dudik2014}.
Quantitatively, inverting the \citet{dz23} Fe\,\textsc{xii}/Fe\,\textsc{xiii}
tables here returns $T_M = 1.47\,\mathrm{MK}$ for a $\kappa = 2.5$ distribution at
$\Teff = 1.5\,\mathrm{MK}$, a $2\%$ offset, well inside the precision of
\citet{lorincik2020}. Any single diagnostic operating
inside the Bethe-weighted ionization moment class is therefore blind to
$\kappa$ at fixed $\langle E\rangle$: it returns a temperature
indistinguishable from $\Teff$ and does not detect the departure from
Maxwellian form. This is the convergence principle of \citet{edmonds2026b},
that every ionization-gated diagnostic structurally returns $\Teff$ regardless
of the source distribution at Knudsen number above $\sim 0.01$. What reveals $\kappa$ is a second diagnostic that projects
the same distribution onto a different moment --- in the corona, the radio
source function of Section~\ref{sec:radio}. The Maxwellian appearance and the
factor-$R$ disagreement are the same fact, seen by one diagnostic and by two.

\section{The transport reading}
\label{sec:transport}

The diagnostic reading takes the distribution as given and asks what two
projections of it return. The transport reading takes the same distribution
and asks what the corona's energy budget does with it. The dominant coronal
loss term is conductive, closed in the standard budget with the
Spitzer--H\"{a}rm form
\begin{equation}
q_\mathrm{SH} = -\kappa_0\,T^{5/2}\,\nabla T ,
\label{eq:spitzer}
\end{equation}
evaluated at the EUV-diagnosed $T \approx \Teff$. Equation~(\ref{eq:spitzer})
is the leading-order Chapman--Enskog closure of the electron heat-flux moment
hierarchy \citep{spitzer1953,braginskii1965}; like every moment closure it
assumes the distribution belongs to a specified family (Maxwellian) and
that the temperature parameter of that family is the moment governing bulk
transport. At $\kappa \approx 2.5$ neither assumption holds.

\subsection{The local closure does not exist}
\label{sec:nonexistence}

Conduction is the third velocity moment, and the Chapman--Enskog integrand
carries the $v^4$ mean-free-path weighting of Coulomb collisions
($\nu \propto v^{-3}$): the conductive flux is carried not by the bulk but by
suprathermal electrons several thermal speeds out. For a Maxwellian this is
harmless: those carriers belong to a population one temperature describes,
the integral converges, and it peaks near $2.3\,v_\mathrm{th}$, which is why
Spitzer--H\"{a}rm is valid for a Maxwellian. For the $\kappa$ distribution it
is not: a hierarchy of moment divergences runs along the $\kappa$ axis
\citep{lazarfichtner2021}: at $\kappa \le 3/2$ the temperature integral
diverges, at $\kappa \le 2$ the heat-flux moment diverges, and at
$\kappa \le 5/2$ the fourth moment governing the next Chapman--Enskog
correction diverges. The standard-$\kappa$ closed-form conductivities carry
poles at $\kappa = 3$ and $\kappa = 4$ \citep{du2013,husidic2021}; the
$e$--$e$-inclusive treatment of \citet{guodu2019} turns negative for
$\kappa < 10$; and the only finite expression, for the regularized $\kappa$
distribution, is cutoff-dependent and of order $10^4$ times the Maxwellian
value \citep{husidic2022}. Across $\kappa \in [2,3]$ there is no finite,
convention-independent, order-unity local conductivity $\kappa_0$ for a
temperature to be substituted into; the finite $-16.5$ (in units of the Maxwellian value) the closed form returns
at $\kappa = 2.5$ is an analytic continuation of the divergent integral, not a
conductivity \citep{edmonds2026c}. It is the Fourier-law form $q = -\kappa_0
\nabla T$ itself that fails, not the finiteness of the physical flux:
structurally the Burnett/super-Burnett divergence of an asymptotic, not
convergent, Chapman--Enskog expansion \citep{grad1949,struchtrup2005}. There
is no local closure to correct at $\kappa = 2.5$; there is only the question
of what the true kinetic flux is.

\subsection{The temperature-substitution trap}
\label{sec:trap}

The result is easy to miss, because the natural way to fold a $\kappa$
correction into the budget is concrete and wrong. If the diagnosed temperature
is the tail-weighted $\Teff$ while the bulk sits at $\Tcore$, the conduction
term looks evaluated at the wrong temperature; substituting $\Tcore$ for
$\Teff$ in (\ref{eq:spitzer}) multiplies the loss by $(\Tcore/\Teff)^{7/2} =
(\kappa/(\kappa-3/2))^{-7/2} \approx 1/25$ at $\kappa = 2.5$, dropping the
standard fluid budget by $\sim 47\%$ \citep{withbroe1977,edmonds2026c}. Every
step is individually valid, and the result is meaningless: the factor corrects
a coefficient that does not exist (Section~\ref{sec:nonexistence}). The
concreteness that makes the correction persuasive is the signature of a
formula evaluated outside its domain: the value extracted depends on where
the divergent integral is truncated, not on the plasma. We carry the $47\%$
not as a result but as an artifact of substituting into a coefficient that
does not exist; we claim no conductive suppression and replace no closure.

\subsection{The deficit, and the flux that remains}
\label{sec:deficit}

The distance from validity is quantitative. The relative entropy of the
$\kappa = 2.5$ distribution from its energy-matched Maxwellian projection is
\begin{equation}
\DKL(\fk \,\|\, \Mxw(\Teff)) = 0.32~\mathrm{nats},
\label{eq:deficit}
\end{equation}
the value of Equation~(\ref{eq:DKL-closed-form}) at $T = \Teff$
(Section~\ref{sec:gap}), derived in closed form in \citet{edmonds2026c}; by the
entropy-deficit
identity \citep{jaynes1957a,csiszar1975,levermore1996} this is exactly the
information the Maxwellian closure discards. It is order unity --- a still
distinctly non-Maxwellian $\kappa = 6$ plasma sits a decade lower at $0.033$
nats --- and model-independent: a projection on the moments of the
distribution, with no collision operator, atomic data, or regularization
entering. It measures how far the Maxwellian closure is from applicable at the
observed $\kappa$, and at $\kappa = 2.5$ that distance is large. This is the $0.32$-nat component of the
partition, met again in Section~\ref{sec:unification}.

The physical heat flux that does exist for $\kappa > 2$ is real but non-local
and tail-dominated. Direct kinetic simulation finds it enhanced over, and able
to reverse sign relative to, the Spitzer--H\"{a}rm value
\citep{landipantellini2001,shoub1983,dorelliscudder1999,dorelliscudder2003}
--- the opposite sense to the trap's apparent suppression --- and the
generalized fluid-moment framework of \citet{cranmerschiff2021} likewise
yields no closed-form $\kappa$-conductivity at $\kappa \approx 2.5$. We do not
compute the flux, and no conclusion here depends on its magnitude or sign. The
adequacy results that do exist --- Spitzer--H\"{a}rm consistent with
$\sim 65\%$ of solar-wind data at $1\,\mathrm{AU}$ for $L_T \gtrsim
3.5\,\lambda_\mathrm{fp}$ \citep{bale2013}, and adequate for $\kappa \gtrsim 5$
\citep{landipantellini2001} --- hold in regimes the quiet-Sun
$\kappa \approx 2.5$ corona is not in; the dissenting view that classical heat
flux survives in strong transition-region gradients \citep{liesvendsen1999} is
the acknowledged counterweight. The claim here is specific to
$\kappa \in [2,3]$, where the published kinetic simulations agree the local
closure fails.

\section{The unification as partition}
\label{sec:unification}

\begin{figure}[htbp]
\centering
\includegraphics[width=0.8\linewidth]{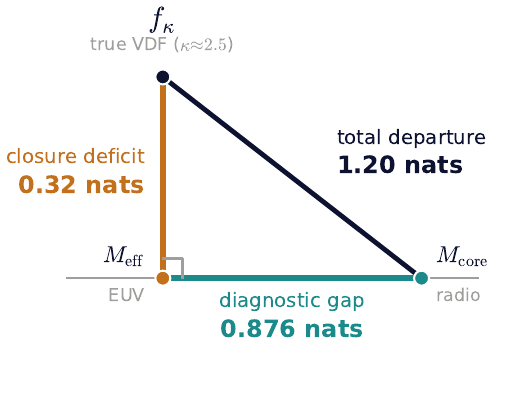}
\caption{The quiet-Sun corona's departure from Maxwellian as one relative
entropy with two orthogonal legs. The $\kappa \approx 2.5$ distribution
$f_\kappa$ lies off the one-parameter Maxwellian family (grey line); its
orthogonal $I$-projection onto that family is the energy-matched Maxwellian
$M_{\mathrm{eff}} \equiv \Mxw(\Teff)$, the temperature the EUV diagnostic
returns, while the radio diagnostic returns a second family member
$M_{\mathrm{core}} \equiv \Mxw(\Tcore)$. The two legs are the closure deficit
$\DKL(\fk \| \Mxw(\Teff)) = 0.32$ nats --- the information the local
Spitzer--H\"{a}rm closure discards --- and the diagnostic gap
$\DKL(\Mxw(\Teff) \| \Mxw(\Tcore)) = 0.876$ nats, a closed-form function of $R$
and the EUV--radio disagreement; the hypotenuse is the total departure
$\DKL(\fk \| \Mxw(\Tcore)) = 1.20$ nats. The right angle marks the
information-geometric orthogonality of the two projections, under which the
cross-term vanishes and the divergences add linearly in nats (the generalized
Pythagorean identity for the $I$-projection; \citealt{csiszar1975}), so
$1.20 = 0.32 + 0.876$. The construction is schematic: the legs are relative
entropies, not Euclidean lengths.}
\label{fig:schematic}
\end{figure}

\begin{figure}[htbp]
\centering
\includegraphics[width=0.8\linewidth]{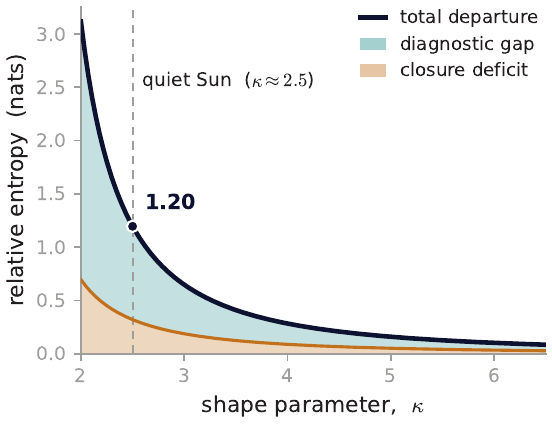}
\caption{The partition of Figure~\ref{fig:schematic} evaluated in closed form
(Equations~\ref{eq:DKL-closed-form} and \ref{eq:gap-identity}) as a function of
the shape parameter $\kappa$. At every $\kappa$ the closure deficit (lower
band) and the diagnostic gap (upper band) stack exactly to the total departure
(curve), and both vanish as $\kappa \to \infty$, where the distribution returns
to Maxwellian. The quiet Sun sits at $\kappa \approx 2.5$, where the legs are
$0.32$ and $0.876$ nats and the total is $1.20$ nats.}
\label{fig:partkappa}
\end{figure}

The two readings have produced two numbers from one distribution: the
diagnostic gap, $0.876$ nats (Section~\ref{sec:gap}), and the closure deficit,
$0.32$ nats (Section~\ref{sec:deficit}). They are not independent. The
Maxwellian $\Mxw(\Teff)$ is the I-projection of $\fk$ onto the one-parameter
Maxwellian family --- the member matching the sufficient-statistic moment
$\langle E\rangle$ and therefore minimizing $\DKL(\fk \,\|\, \cdot)$ over the
family (Section~\ref{sec:projection}). For an I-projection onto an exponential
family, the generalized Pythagorean theorem for Bregman divergences gives, for
any other family member $\Mxw(T)$,
\begin{equation}
\DKL(\fk \,\|\, \Mxw(T))
= \DKL(\fk \,\|\, \Mxw(\Teff)) + \DKL(\Mxw(\Teff) \,\|\, \Mxw(T)) ,
\label{eq:pythagoras}
\end{equation}
with no cross-term \citep{csiszar1975,covthomas2006}. Setting $T = \Tcore$,
\begin{equation}
\underbrace{1.19523}_{\DKL(\fk\|\Mxw(\Tcore))}
\approx \underbrace{0.31967}_{\text{closure deficit}}
+ \underbrace{0.87556}_{\text{diagnostic gap}} \quad \mathrm{nats},
\label{eq:partition}
\end{equation}
identity~(\ref{eq:pythagoras}) is exact; the displayed values are rounded. The closure deficit and the
diagnostic gap are the two orthogonal components of one divergence: the distance from $\fk$ to its best Maxwellian, and the distance
from that best Maxwellian to the one the radio diagnostic returns. They account for
the total departure from the radio-diagnostic Maxwellian exactly, with no
remainder; the partition is shown schematically in
Figure~\ref{fig:schematic} and evaluated across $\kappa$ in
Figure~\ref{fig:partkappa}. This is the sense in which closure failure and
diagnostic disagreement are the same relative entropy.

\emph{Same object, not equal scalars.} The two legs are unequal ($0.32 \neq
0.876$), and the claim is not that they are. It is that they are the two
legs of one right triangle in the geometry of the divergence: the closure
deficit is the leg from the true distribution to the foot of its projection,
the diagnostic gap is the leg from that foot to the radio Maxwellian, and the
$1.20$-nat total is the divergence they compose. Orthogonality is what makes
the decomposition additive rather than a relabeling: because $\Mxw(\Teff)$ is
the I-projection, the cross-term vanishes identically, and the two references
against which the legs are measured are nested projections of one source onto
the Maxwellian family and then onto a specific member of it.

The mean-energy (D\&D) convention is not a free choice here: the EUV diagnostic
reads the energy moment (Section~\ref{sec:euv}), the sufficient statistic of the
Maxwellian family, so holding $\langle E\rangle$ fixed across $\kappa$ is what
makes $\Mxw(\Teff)$ exactly the energy-moment match and hence the I-projection on
which (\ref{eq:pythagoras}) rests; the values in (\ref{eq:partition}) are
evaluated at that ideal projection ($R_0 = 2.500$). The physical EUV diagnostic
realizes the energy moment only to within $\varepsilon_\mathrm{EUV}$
(Section~\ref{sec:euv}), so the observed $R \approx 2.4$ is an accurate proxy
for the exact triangle, not a departure from it. Because the I-projection
minimizes the divergence, the deficit leg is second-order insensitive to the
residual ($\sim 10^{-4}$ nats at $\varepsilon_\mathrm{EUV} = 0.013$); the
diagnostic gap and the additivity defect inherit $\varepsilon_\mathrm{EUV}$ at
first order, $\approx \tfrac{3}{2}(R_0 - 1)\,\varepsilon_\mathrm{EUV} \approx
0.03$--$0.04$ nats at the line-ratio values, bounded by the $0.71$--$0.90$-nat
envelope already carried (Section~\ref{sec:gap}) and small against the
$1.20$-nat total.

The entropy-deficit reading of a moment closure is not itself new:
\citet{levermore1996} established that a moment closure is an entropy
minimization subject to moment constraints, discarding the relative entropy
between the true distribution and its moment-matched projection, and
\citet{abdelmalik2016} extended closures from relative entropy to the broader
$\varphi$-divergence family. The present decomposition sits in the Bregman
branch of that same generalization: the Kullback--Leibler and Itakura--Saito
divergences it uses are Bregman divergences, the two families meeting at the
relative-entropy member. The identification is the new step: that the relative entropy a
coronal transport closure discards
and the relative entropy two coronal diagnostics disagree by are the two
orthogonal coordinates of a single divergence of the electron distribution
from Maxwellian. The partition is a theorem about derived quantities.

\section{The generator}
\label{sec:generator}

The partition of Section~\ref{sec:unification} fixes a number and leaves a
question. The corona's departure from Maxwellian is one relative entropy,
$1.20$ nats, accounted for without remainder by the diagnostic gap ($0.876$ nats) and
the closure deficit ($0.32$ nats); both components are read off a single shape
parameter, $\kappa \approx 2.5$, stable across the solar cycle
\citep{mercierchambe2015,edmonds2026a}. What the partition does not say is what
sets that $\kappa$. The flux that carries the departure is non-local and
tail-dominated: across $\kappa \in [2,3]$ the local Spitzer--H\"{a}rm closure
has no convergent form, and the finite value it returns at $\kappa = 2.5$ is
an analytic continuation of a divergent integral, not a conductivity
\citep{edmonds2026c,du2013,husidic2021}. The mean free path scales as $v^4$, so
the suprathermal population is effectively collisionless even where the bulk is
marginally so \citep{shoub1983}, and the bulk Knudsen number $\Kn \sim
0.01$--$0.1$ places the whole structure in the kinetic regime
\citep{landipantellini2001,dorelliscudder1999,edmonds2026a}. The physical flux
that remains is real but non-local: it is the Fourier-law form itself, not the
finiteness of the flux, that fails \citep{edmonds2026c}. We do not solve for
the generator here, and we take no side on the magnitude or sign of that flux.
We read off, from the structure of the entropy itself, the conditions any
generator must meet, and ask which mechanisms in the literature meet them.

\subsection{Three constraints}
\label{sec:constraints}

The constraints are consequences of how the two components are built, not a wishlist
assembled to favor a conclusion.

\paragraph{(a) Tail-acting.} The closure deficit $\DKL(\fk \,\|\, \Mxw(\Teff))
= 0.32$ nats is the information the energy-matched Maxwellian discards: by the
entropy-deficit identity it is the gap $H(\Mxw) - H(\fk)$ between the two, a
definitional measure of how far the closure is from applicable
\citep{edmonds2026c,levermore1996,csiszar1975}. The mismatch lives in the
shape, not the bulk --- the energy-matched Maxwellian is simultaneously too hot
for the $\kappa$ core and too thin in its tail --- and the physics the deficit
governs, conduction, reads that mismatch through its suprathermal end: the
conductive moment carries the $v^4$ mean-free-path weighting of Coulomb
collisions, so it is set by electrons several thermal speeds out, not by the
bulk \citep{edmonds2026c,shoub1983}. The part of the distribution that both
makes the deficit nonzero and carries the flux is the tail. A mechanism that
reshapes only the thermal core, leaving the tail Maxwellian, produces neither.
The generator must act on the suprathermal tail.

\paragraph{(b) Non-local.} The tail that carries the deficit is, at the
energies that carry it, collisionless. With $\lambda_\mathrm{mfp} \propto v^4$
\citep{shoub1983}, the threshold above which electrons are collisionless over
the gradient scale sits at $v_\mathrm{crit}/v_\mathrm{th} \approx 2.1$
($E \gtrsim 4.5\,k\Teff$) for the bulk Knudsen number $\Kn \sim 0.05$; at
$\kappa = 2.5$ this collisionless tail is some $5\%$ of the electrons but holds
about $40\%$ of the kinetic energy \citep[computed in][]{edmonds2026c}. That population's
occupation at a point is set not by a local rate but by the distribution
streaming in along the field from a scale height away. A quantity fixed by
conditions a scale height removed cannot be set by a local source term. This is
the same non-locality that makes the local closure non-existent: the flux and
the deficit are two readings of one collisionless tail. The generator must set
$\kappa$ non-locally. This is the load-bearing constraint, and it carries the
weight on Knudsen-number grounds: a mechanism local in configuration space acts
through a production rate at a point, and a collisionless tail does not reach
steady state under a local rate.

\paragraph{(c) $\kappa$-fixing.} The partition is a partition of one $\kappa$.
Both components are functions of $\kappa \approx 2.5$, and the eight-year stability
of $R$ is the stability of that single value \citep{mercierchambe2015}. A
mechanism that merely permits suprathermal tails --- that admits a family of
$\kappa$ without selecting one --- leaves the observed band unexplained. The
generator must fix $\kappa$ to the observed band, not license a range that
contains it.

\subsection{The candidates against the constraints}
\label{sec:contest}

Read against (a)--(c), the candidate generators in the literature separate
cleanly (Table~\ref{tab:rivals}). Each produces a suprathermal tail, so each
clears (a). They divide on (b).

\begin{table}[t]
\centering
\caption{Candidate $\kappa$-generators against the three constraints.}
\label{tab:rivals}
\begin{tabular}{lccc}
\toprule
Mechanism & (a) tail & (b) non-local & (c) fixes $\kappa$ \\
\midrule
Stochastic / weak-turbulence$^{a}$ & yes & no & special closure only \\
Collisionless relaxation$^{b}$ & yes & no & not gradient-tied \\
Reconnection / nanoflares$^{c}$ & yes & partial & no \\
Whistler / wave heating$^{d}$ & yes & partial & no \\
Velocity filtration$^{e}$ & yes & yes & open \\
\bottomrule
\end{tabular}

\smallskip
{\footnotesize\noindent
$^{a}$~\citet{yoon2012,yoon2014};\quad
$^{b}$~\citet{ewart2025};\quad
$^{c}$~\citet{che2014};\quad
$^{d}$~\citet{vocks2008};\quad
$^{e}$~\citet{scudder1992a,scudder1992b}.}
\end{table}

Two of the candidates are local in configuration space. Stochastic and
weak-turbulence acceleration sets the tail by velocity-space diffusion at a
point \citep{yoon2012,yoon2014}; it is the nearest miss on (c), since
self-consistent Langmuir turbulence selects $\kappa = 9/4$, close to the
observed band, but it selects it through a process local in configuration space
and not tied to the gradient that carries the flux. Collisionless relaxation
yields a universal power-law tail from velocity-space diffusion
\citep{ewart2025}, a result we credit, but the tail it fixes is set by the
relaxation, not the transport, and is likewise not gradient-tied. Both fail (b)
as standalone steady-state generators, since a collisionless tail does not reach
steady state under a rate local in configuration space; neither is thereby
excluded as the local-replenishment half of the non-local transport hybrid that
constraint (c) leaves open (Section~\ref{sec:open}).

Two further candidates are not simply local, and must be distinguished rather
than dismissed. Whistler-driven wave heating produces coronal suprathermal tails
from an initially Maxwellian plasma, and the waves propagate
\citep{vocks2008,vocks2016}; it is the competing \emph{generative} account, a
tail grown in place by wave--particle interaction rather than a tail surviving
non-local transport, and it does not fix $\kappa$ to the band. Reconnection and
nanoflare heating likewise drive non-local transport, through beams and
conduction fronts along the field \citep{che2014}, but their time-averaged
profiles do not self-consistently select the observed steady quiet-Sun
$\kappa$, which is incidental to the heating rather than fixed by it. What (b) requires is the non-local supply of the ambient tail itself, not of
the energy or beams that produce it; that distinction, not a denial that waves
or beams propagate, is what leaves each only partway into (b) and stops at (c). Both remain live,
however, as the local generation of a tail that non-local transport then
redistributes, one of the two complementary possibilities constraint (c) leaves
undiscriminated (Section~\ref{sec:open}); neither fixes $\kappa$ on its own.

One mechanism clears (a) and (b) together as a standalone account. Velocity filtration is, by
definition, a non-local transport statement: fast electrons cross a potential
gradient because their kinetic energy exceeds the barrier, so the suprathermal
tail at a point is set by the population admitted from below, not produced in
place \citep{scudder1992a,scudder1992b}. Its tail is the collisionless tail of
constraint (b), and the Knudsen number that makes the closure non-existent is
the same number that lets filtration operate \citep{edmonds2026a}. It alone
among the candidates satisfies the non-locality that does the discriminating.
The comparison therefore reduces to a single constraint, (c): whether
filtration reaches it.

\subsection{The open constraint}
\label{sec:open}

Constraint (c) is open, and we leave it open. The published record on
filtration states what it establishes and no more: filtration permits and
redistributes a suprathermal tail set as a boundary condition at the base; it
does not, in any published treatment, fix $\kappa$ to a value from first
principles. \citet{anderson1994} showed that below $\sim 0.5\,R_\odot$ the
collisional corrections to the collisionless filtration picture are of order
unity or larger, so a purely collisionless derivation of the tail is
incomplete exactly where the quiet-Sun base sits; his conclusion is that the
picture must be redone with collisions included self-consistently.
\citet{landipantellini2001} found that, unless a hard tail is imposed at the
base, filtration alone does not sustain the gradient: the tail is
redistributed, not generated. \citet{scudder2019} ties the degree of
non-thermality to a single dimensionless measure of the parallel electric
field, which relocates (c) to the value of that field at the quiet-Sun base
rather than closing it. Whether the tail is generated locally and then
redistributed non-locally, or set non-locally outright, is itself part of what
(c) leaves unsettled, and we do not settle it here: local replenishment of
the tail against Coulomb relaxation and non-local supply along the field are
complementary and, on present evidence, undiscriminated \citep{edmonds2026c}.
First-principles filtration models with intermittent chromospheric heating are
the live front on exactly this question \citep{barbieri2024,barbieridemoulin2025};
they are in progress, not a settled derivation of a fixed $\kappa$.

The target band, moreover, is marked out before any generator is named, from a
direction that owes nothing to the diagnostic ratio. The local closure has no
usable convergent form anywhere below the conductivity poles at $\kappa = 3$ and
$\kappa = 4$; Spitzer--H\"{a}rm recovers only for $\kappa \gtrsim 5$
\citep{landipantellini2001}, and $\kappa \approx 2.5$ sits well inside that
sub-pole regime. The diagnostics fix the value; the transport, separately and
from the pole structure alone, marks that sub-pole regime as the one where
no local closure exists. This does not locate $\kappa \approx 2.5$ (the
diagnostics do that alone), but it removes the regime from contention: the open
question is not where a generator must land, which the diagnostic reading already
places squarely in the closure-failure window, but whether non-local transport
has a fixed point there.

The constraint structure is the result. Three conditions follow from how the
entropy is built; the candidate generators reduce to the one
mechanism whose definition is a non-local transport statement; and whether that
mechanism, or any mechanism, fixes $\kappa$ to the observed band is the single
thing the partition does not decide.

\section{Conclusions}
\label{sec:conclusion}

The quiet-Sun coronal departure from Maxwellian is one relative entropy. Its
two orthogonal components --- the $0.876$-nat gap two diagnostics disagree by, and
the $0.32$-nat deficit the local conductive closure discards --- are the
diagnostic disagreement and the closure failure, and the generalized Pythagorean
identity makes them the same divergence object in two coordinates. The partition is, in
effect, a $\kappa$-gauge: the same shape parameter is read by the diagnostics
and by the transport, and the eight-year stability of $R$ is the stability of
that single projection structure rather than of any fluctuating kinetic
quantity. What sets the gauge --- what fixes $\kappa$ to the observed band ---
is the constraint the partition poses and does not close.

\section*{Acknowledgments}
The author thanks the Dzif\v{c}\'{a}kov\'{a}--Dud\'{\i}k group at Ond\v{r}ejov
for the CHIANTI-compatible kappa rate tables that make the diagnostic reading
computable. Computational verification of the analytical results was assisted
by Claude (Anthropic); all scientific arguments and conclusions are the
author's own.

\medskip
\noindent\textit{Data availability.} The results reported here are analytical
and reproducible in full from the closed-form expressions of
Sections~\ref{sec:diagnostic}--\ref{sec:unification}
(Equations~\ref{eq:R-closed-form}, \ref{eq:DKL-closed-form},
\ref{eq:gap-identity}, and \ref{eq:pythagoras}); no new observational data were
generated. The $\kappa$-resolved and matched Maxwellian ion-equilibrium values
quoted in Section~\ref{sec:numerical} are taken from the publicly available
KAPPA database \citep{dz23} on the CHIANTI v10.1 atomic-data basis.

\medskip
\noindent\textit{Software.} NumPy, SciPy, mpmath.

% =============================================================================
% References
% =============================================================================

\end{document}